\documentclass[preprint]{sig-alternate-per}
\usepackage{epsfig,amsmath,amsfonts}
\usepackage{url}

\graphicspath{{.}{./figures/}}

\begin{document}

\CopyrightYear{2014}

\pagenumbering{arabic}

\title{Polynomial Histograms for Memory-Efficient Representation of
  Long-tailed System Distributions}

\subtitle{2014 Preprint}

\author{
  \alignauthor
  Murray Stokely,
  Tim Hesterberg,
  Nate Coehlo,
  Arif Merchant\\
  \affaddr{Google, Inc.} \\
  \email{\small\tt\{mstokely,rocket,natec,aamerchant\}@google.com}
}
\maketitle

\begin{abstract}

Distributed systems must frequently keep track of many different types
of performance metrics across
many
different computers.  For example, the latency distribution of certain
operations may be computed for a large combination of
computers, users, and operations.  These
empirical distributions need to be collected at minimal expense on
the individual software components, efficiently aggregated across
multiple dimensions, and stored in a compact representation for a variety of
downstream data analysis applications.

We describe an information loss metric for binned data that allows us
to optimize cost of information loss from different histogram
representations.
We explore the use of polynomial
histograms where each bin of a histogram is annotated with moments of
the underlying distribution in that bin.  These polynomial histograms
are compared to traditional histograms using the same storage
cost for additional bins instead of annotations in each bin.  We
describe an application of these techniques for file system metrics
for a large production
system, and analytically
characterize when polynomial histograms offer more
information at lower cost.

\end{abstract}

\section{Introduction}

In many cloud scale systems, monitoring, measuring, and logging
performance metrics is extremely difficult because there are literally
billions of possibly interesting metrics. For example, in a large
distributed file system, we may wish to monitor the resource usage,
throughput, and latency per active user so that we can track down
sources of performance anomalies~\cite{tailscale}. To detect
interactions with other applications and further narrow down the
source of performance issues, we may wish to monitor such metrics for
combinations of users and resources.  However, the memory requirements
for comprehensive logging at this scale are usually exorbitant. An
alternative is to maintain aggregate statistics, such as the empirical
distributions of the metrics, as histograms. Even so, the memory
requirements for maintaining histograms for a large number of metrics
can be burdensome; as such, it is critical to make the histograms as
efficient as possible to minimize information loss while limiting the
memory used.


In cloud data centers, it is common to collect histograms per user,
per server, for a variety of metrics - IO delays, network latencies,
CPU throttling, etc. - so that Service Level Agreements can be
monitored and resources allocated appropriately.  For example,
Google's Janus system for flash provisioning~\cite{janus}
collects histograms for the ages of files read and also the age of files
stored, per workload group, which can be individual users, subsets of
files from each user, or even groups of columns from a user's
tables. Since there can be tens of thousands of servers, thousands of
users, and tens of metrics being monitored per user/server, there can
be billions of histograms maintained in a datacenter.

Memory requirements for histograms can be reduced in several ways.
We can use coarser bins, but this reduces the fidelity of the histogram. We
could dynamically adjust bin boundaries to improve accuracy, but
this requires additional processing on sensitive nodes, may introduce
non-deterministic overhead, and makes aggregation difficult or impossible across
computers.


We explore the effectiveness of using {\em polynomial
  histograms,} where the number of bins is reduced, but the
distribution of samples within each bin is maintained using a
low-order polynomial. For a fixed amount of memory, when is it
preferable to store polynomial annotations in coarser bins?

Our contributions are as follows: (1) We compare the information loss
due to normal (fixed bin) histograms to those with a moment annotation;
(2) we compare the errors empirically for some empirical distributions of system metrics
in a cloud environment; and (3) we give rules of thumb for when
using polynomial histograms is effective.


\section{Background}

A variety of techniques have been developed to make synopses of
massive data sets.  Streaming quantile algorithms
\cite{chambers2006monitoring} keep an approximation of a given
quantile of the observed values in a stream.  These algorithms are
most useful when there are a small number of quantiles of interest,
but they do not offer a density estimate across the full distribution
for cases where a variety of downstream data analysis will be done
based on the synopsis.

There has been a lot of work in the database community on histograms
that dynamically adjust bin breakpoints as new data are seen to
minimize error, but these methods are less useful for distributions
with a small number of samples, and the resulting irregular bins are
harder to combine as part of a distributed computation.

Information loss metrics for fixed-boundary histograms of file system parameters are
explored in \cite{douceur1999large}. We utilize similar information
loss metrics but also consider the space versus information loss
tradeoff of adding additional moments to each bin of the histogram to
build low-order ``Polynomial'' \cite{sagae1997bin} or ``Spline''
\cite{Poosala:1997:HET:269157} histograms.



As with the work of K{\"o}nig and Weikum \cite{konig1999combining} we
do not require continuity across bucket boundries, and
find that this attribute is essential in order to accurately capture
large jumps in bucket frequencies.  Unlike that work, however, we are
not focused on optimal dynamic partitioning of bucket boundaries and
our information loss metric is focused around making definitive statements.
Instead, we are
focused on constrained resource environments where the computational
and memory requirements of those techniques would be excessive.



\subsection{Information Loss Due to Binning}

Binning of an empirical
distribution into a histogram representation introduces a form of
preprocessing that constrains all later analyses based on that data
\cite{blocker2013potential}.  Bin breakpoints are often fixed in
advance for specific system quantities to reduce the computational
overhead of keeping track of many different histograms.
However, bin breakpoints that are
poorly chosen relative to the underlying data set may introduce
considerable error when one tries to compute means or percentiles
based on the histogram representations.  This is especially true for
the exponentially bucketed (e.g. buckets that double in size)
representations of distributions such as
latencies or arrival times that have a large dynamic range.


%

In evaluating representations of system distributions, we define the
\emph{Earth Mover's Distance of the Cumulative Curves} (EMDCC) as our information
loss metric.  In particular, if $X$ is the (unknown) underlying data set with
distribution $F$, and $h$ is our data representation, r is the range of our representation,
then we define upper and lower bounds $F_{h+}(x) \quad F_{h-}(x)$ as the highest and lowest possible
values for the true distribution given the observed representation, and the
EMDCC as the normalized L1 distance between them:

$$ EMDCC(X, h) = \frac{1}{r} \int_{\mathbb{R}} |F_{h+}(x) - F_{h-}(x)|dx $$

\noindent noting that in the case that $h$ is a histogram bucketing scheme,
$F_{h-}$ always puts its mass on the left endpoints and $F_{h+}$ always puts
near the right endpoints.  The Earth Mover's Distance
\cite{rubner2000earth} is also known as the
Mallows distance, or Wasserstein distance with $p=1$ in
different communities.

\begin{figure}[h!]
\centering
\includegraphics[width=\linewidth]{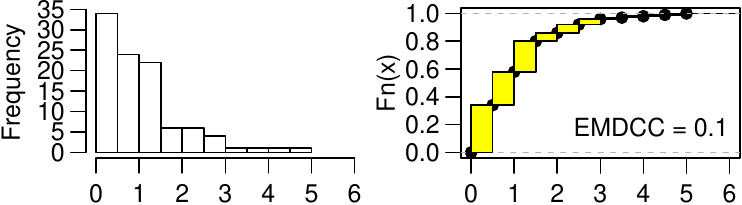}
\caption{An example histogram (left) with its CDF representation and a
  yellow area of uncertainty showing where the true empirical cdf of
  the unbinned data must lie (right).}
\label{fig:emdcc}
\end{figure}

Figure~\ref{fig:emdcc} shows a histogram (left) along with the CDF representation and the
associated area of uncertainty (in yellow).  Any underlying dataset
having the given histogram representation must have a true ecdf lying
entirely within the yellow area.  A histogram with more granular
buckets would reduce the information loss at the expense of additional
storage space to store the buckets.

Adding more buckets as in this example usually reduces the EMDCC, but
are there more efficient ways to reduce the EMDCC for a given amount
of storage space?

\subsection{Polynomial Histograms}

Given a fixed amount of storage space, we can trade off the granularity of
histogram buckets for additional statistics within each bucket.  For example,
in addition to storing the counts between histogram boundaries $(a, b)$, we
could also store the the mean and higher moments.  Histograms with
annotations of moments per bin are known as \emph{Polynomial
Histograms} \cite{sagae1997bin}.  Storing the moments is appealing in
a distributed systems context because merging histograms with the same
bucket boundaries remains trivial.  We will use the notation $H(b,p)$ to
denote a histogram with $b$ bins annotated with the $p$-moments of
each bin.

Knowing the first moment can help a lot when it is near the boundary; the
EMDCC associated with bucket $(a, b)$ will be zero if the mean is $a$.
In general, with many points in bucket, a continuous approximation says that a mean of $\mu = \alpha * a + (1-\alpha) * b$ gives an EMDCC of

$$
\lambda(\alpha) = \alpha * ln(\frac{1}{\alpha}) + (1 - \alpha) * ln(\frac{1}{1-\alpha})
$$

The function $\lambda(\alpha)$ is symmetric around $0.5$, is increasing up to it's max of $\lambda(0.5) \approx 0.7$, integrates to $0.5$, and $\lambda(0.2) = \lambda(0.8) \approx 0.5$.  Since bisection always halves the EMDCC, this gives rules of thumb about the merits of bisection vs. storing the mean; if $\alpha < 0.2$ or $\alpha > 0.8$ then storing the mean is better, storing the mean can be worse than bisection by 40\% but it can also be infinitely better, if $\alpha's$ are uniformly distributed and independent of the counts per bucket then bisection and storing the mean should give the same reduction on average.  If the true density is smooth enough relative to the bucketing scheme, then $\alpha$ will tend to be closer to $\frac{1}{2}$, which implies inferiority of keeping the mean with respect to the EMDCC metric.

\noindent
\textbf{Proof:} \\

We restrict our attention to $(a,b)$ where we also know the $p^{th}$ moment$\big(\mu_p^p = \frac{1}{n} \sum_{i=1}^n x_i^p \big)$.  We construct $F_{h+}(x), F_{h-}(x)$ pointwise as the upper and lower bound curves, then integrate to find the reduction in EMDCC. 

For $x \in (a, \mu_p)$, the lower bound is $F(a)$ and the upper bound has support $\{x, b\}$.  This implies that $\mu_p^p$ equals

$$
\frac{F_{h+}(x)-F(a)}{F(b)-F(a)} * x^p + \frac{F(b)-F_{h+}(x)}{F(b)-F(a)} * b^p
$$

Therefore,

$$
F_{h+}(x) = F(a) + (F(b)-F(a)) \frac{b^p-\mu^p}{b^p-x^p}
$$

For $x \in (\mu_p, b)$, the upper bound is $F(b)$ and the lower bound has support $\{a+\epsilon, x+\epsilon\}$ where we let $\epsilon \rightarrow 0$.  This implies that $\mu_p^p$ equals

$$
\frac{F_{h-}(a+\epsilon)-F(a)}{F(b)-F(a)} * (a+\epsilon)^p + \frac{F_{h-}(x+\epsilon)-F_{h-}(a + \epsilon)}{F(b)-F(a)} (x + \epsilon)^p
$$

Therefore, re-arranging, noting that $F_{h-}(a+\epsilon)=F_{h-}(x)$, $F_{h-}(x+\epsilon)=F(b)$, and letting $\epsilon \rightarrow 0$ gives

$$
F_{h-}(x) = F(b) - (F(b)-F(a)) \frac{\mu_p^p - a^p}{x^p-a^p}
$$

Next, the area reduction from knowing the moment comes from the integral between upper and lower bounds:

$$
\frac{1}{(F(b)-F(a))(b-a)}\int_{a}^{b} |F_{h+}(x) - F_{h-}(x)|dx =
$$
$$
\frac{1}{b-a} \bigg(
\int_{a}^{\mu_p} \frac{b^p-\mu_p^p}{b^p-x^p}dx + \int_{\mu_p}^{b} \frac{\mu_p^p-a^p}{x^p-a^p}dx
\bigg)
$$

\noindent
and this gives the stated result when $p=1$.  More complex formulas exist when multiple moments are known simultaneously.

\begin{figure}[h!]
\centering
\includegraphics[width=\linewidth]{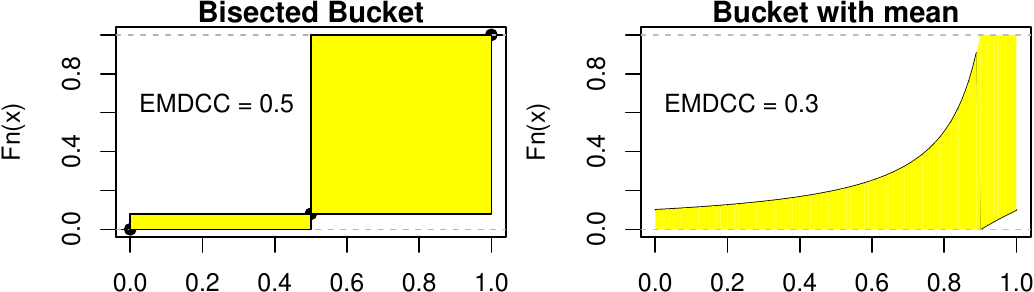}
\caption{Yellow areas of uncertainty for where the ecdf of the
  unbinned data must lie given a histogram bin bisected in two (left)
  or a histogram annotated with the mean of values in that bin (right).}
\label{fig:polyemdcc}
\end{figure}

Figure~\ref{fig:polyemdcc} uses an example bin with points taken from
Beta(0.5, 0.05) and a mean value of 0.9 to illustrate visually the
tradeoff in information loss between $H(2,0)$ and $H(1,1)$ histograms.
Knowing the mean value in this case
constrains the area where an ecdf of the underlying distribution with
that binned representation lies more than if we had just added twice
as many bins at the same storage cost.

\section{Empirical Validation}

We test the efficacy of polynomial histograms in tracking read sizes
for 315 storage users in one of
 Google's
data centers.  For this
system, we are interested in log read sizes, and we restrict our range from $log(0) = 1 \text{byte}$ to $log(24) = 16\text{MB}$ and find that storing mean and counts in each of the 24 buckets is far more effective than bisecting into 48 buckets.

If we do not store the mean, then K equally sized bins will give an EMDCC of 1/K.  When we store the mean in K equally sized bins and get an EMDCC of X, then we define
$$ \text{information gain} \quad = \frac{1}{2*K*X} $$
A value of 5 implies we would need 5 times as much storage space from equally spaced buckets to achieve the the same EMDCC.  This is bounded below by 1/1.4=0.73, but can get arbitrarily large.

\begin{figure}[h!]
\centering
\includegraphics[width=\linewidth]{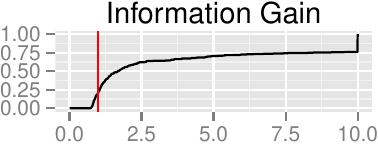}
\caption{The Information Gain from storing the mean in 24 integer buckets of log file sizes across 315 storage users.}
\label{fig:validation}
\end{figure}

Figure ~\ref{fig:validation} shows the information gain associated with 24 integer buckets of log file sizes, where we have truncated gains above 10.  While $\approx20\%$ see a minor loss, approximately the same number see gains over 10, and 40\% see gains over 2.5.  This shows that 24 buckets with means is superior to 48 regular buckets, and we have also checked that 12 buckets with means is superior to 24 regular buckets, although the relative gain is slightly smaller.  On the other hand, 6 buckets with means is slightly worse 12 regular buckets, because the biggest discontinuities are less likely to sit on the endpoints at this scale.

%

\section{Conclusion}
\label{sec:Conclusions}

Cloud data centers monitor a very large number of metric distributions,
particularly for latency metrics, as compact histograms. Memory for
these histograms is limited, and so it is important to use a
representation that minimizes information loss without increasing the
memory footprint. We describe an information loss metric for
histograms, and show that, by using histograms with fewer bins but
adding information about the moments of the samples in the bin,
information loss can be reduced for certain classes of distributions,
and that such distributions occur commonly in practice.
An open-source R package for analyzing the information loss due to
\cite{histogramtoolsredacted}.
The package includes example code to
generate all figures included in this paper, and the data set
used in the empirical validation section.

%


{\footnotesize \bibliographystyle{abbrv}
\bibliography{refs}}

@Manual{histogramtoolsredacted,
    title = {Redacted Package},
    author = {Redacted Author},
    year = {2013},
    note = {R package version Redacted},
    url = {https://redacted.com},
  }

@article{blocker2013potential,
  title={The potential and perils of preprocessing: Building new foundations},
  author={Blocker, Alexander W and Meng, Xiao-Li},
  journal={Bernoulli},
  volume={19},
  number={4},
  pages={1176--1211},
  year={2013},
  publisher={Bernoulli Society for Mathematical Statistics and Probability}
}

@article{rubner2000earth,
  title={The earth mover's distance as a metric for image retrieval},
  author={Rubner, Yossi and Tomasi, Carlo and Guibas, Leonidas J},
  journal={International Journal of Computer Vision},
  volume={40},
  number={2},
  pages={99--121},
  year={2000},
  publisher={Kluwer Academic Publishers}
}

@article{sagae1997bin,
  title={Bin interval method of locally adaptive nonparametric density
estimation},
  author={Sagae, Masahiko and Scott, DW},
  journal={Statistics technical report of RICE University},
  pages={1--21},
  year={1997}
}

@article{douceur1999large,
  title={A large-scale study of file-system contents},
  author={Douceur, John R and Bolosky, William J},
  journal={ACM SIGMETRICS Performance Evaluation Review},
  volume={27},
  number={1},
  pages={59--70},
  year={1999},
  publisher={ACM}
}

@article{chambers2006monitoring,
  title={Monitoring networked applications with incremental quantile estimation},
  author={Chambers, John M and others},
  journal={Statistical Science},
  pages={463--475},
  year={2006},
  publisher={JSTOR}
}

@phdthesis{Poosala:1997:HET:269157,
 author = {Poosala, Viswanath},
 title = {Histogram-based Estimation Techniques in Database Systems},
 year = {1997},
 note = {UMI Order No. GAX97-16074},
 publisher = {University of Wisconsin at Madison},
 address = {Madison, WI, USA},
}

@inproceedings{konig1999combining,
  title={Combining histograms and parametric curve fitting for feedback-driven query result-size estimation},
  author={K{\"o}nig, Arnd Christian and Weikum, Gerhard},
  booktitle={VLDB},
  pages={423--434},
  year={1999},
  organization={Morgan Kaufmann Publishers Inc.}
}

@inproceedings{janus,
  title={Janus: optimal flash provisioning for cloud storage workloads},
  author={Albrecht, Christoph and others},
  booktitle={Proceedings of the USENIX ATC},
  pages={91--102},
  year={2013},
  organization={USENIX Association}
}

@article{tailscale,
 author = {Dean, Jeffrey and Barroso, Luiz Andr{\'e}},
 title = {The Tail at Scale},
 journal = {Communications of the ACM},
 issue_date = {February 2013},
 volume = {56},
 number = {2},
 month = feb,
 year = {2013},
 issn = {0001-0782},
 pages = {74--80},
 numpages = {7},
 url = {http://doi.acm.org/10.1145/2408776.2408794},
 doi = {10.1145/2408776.2408794},
 acmid = {2408794},
 publisher = {ACM},
 address = {New York, NY, USA},
}


\end{document}